\documentclass[twocolumn,pra,superscriptaddress,floatfix,showpacs,nofootinbib]{revtex4}
\usepackage[dvips]{epsfig}
\usepackage{graphicx} 
\usepackage{bm} 
\usepackage{braket}
\usepackage[intlimits]{amsmath}
\usepackage{amsfonts,amssymb,amstext,amscd}
\usepackage{color}
\usepackage{hyperref}
\usepackage{subfigure}
\usepackage{bbm}

\newcommand{\ns}[1]{\mkern-#1mu}
\newcommand{\bs}[1]{\boldsymbol{#1}}
\newcommand{\re}{\operatorname{Re}}
\newcommand{\im}{\operatorname{Im}}
\newcommand{\tr}[1]{\operatorname{Tr}\big[#1\big]}
\newcommand{\bracket}[3]{\langle#1|#2|#3\rangle}
\newcommand{\ketbra}[2]{|#1\rangle\mkern-1mu\langle#2|}
\renewcommand{\bra}[1]{\left\langle#1\right|}
\renewcommand{\ket}[1]{\left|#1\right\rangle}
\renewcommand{\braket}[2]{\langle#1|#2\rangle}

\graphicspath{{./},{./Figures/}}

\begin{document}

\title{Vortex formation and dynamics in two-dimensional driven-dissipative condensates} 

\author{F. Hebenstreit}
\email{hebenstreit@itp.unibe.ch}
\affiliation{Albert Einstein Center, Institut f\"{u}r Theoretische Physik, Universit\"{a}t Bern, 3012 Bern, Switzerland}

\begin{abstract}
 We investigate the real-time evolution of lattice bosons in two spatial dimensions whose dynamics is governed by a Markovian quantum master equation.
 We employ the Wigner-Weyl phase space quantization and derive the functional integral for open quantum many-body systems that governs the time evolution of the Wigner function.
 Using the truncated Wigner approximation, in which quantum fluctuations are only taken into account in the initial state whereas the dynamics is governed by classical evolution equations, we study the buildup of long-range correlations due to the action of non-Hermitean quantum jump operators that constitute a mechanism for dissipative cooling.
 Starting from an initially disordered state corresponding to a vortex condensate, the dissipative process results in the annihilation of vortex-antivortex pairs and the establishment of quasi long-range order at late times.
 We observe that a finite vortex density survives the cooling process which disagrees with the analytically constructed vortex-free Bose-Einstein condensate at asymptotic times.
 This indicates that quantum fluctuations beyond the truncated Wigner approximation need to be included to fully capture the physics of dissipative Bose-Einstein condensation.
\end{abstract}


\maketitle

\section{Introduction}\vspace{-0.3cm}
Recent advances in atomic, molecular and optical (AMO) physics have made it possible to experimentally study equilibrium and nonequilibrium quantum many-body systems under controlled laboratory conditions \mbox{\cite{Jane:2003,Bloch:2005,Bloch:2012}}.
In this respect, open quantum systems have attracted much interest in recent years as they provide a resource for quantum computation \cite{Verstraete:2009,Sinayskiy:2014}, quantum simulation \cite{Sweke:2015,Zanardi:2016} and quantum state preparation \cite{Verstraete:2009,DallaTorre:2013,Caballar:2014}.
For instance, tailored couplings to an environment have been suggested to generate strongly correlated states such as condensates of bosons \cite{Diehl:2008}, d-wave pairing states or antiferromagnetic states of fermions \cite{Diehl:2010,Weimer:2016}, or topologically nontrivial states \cite{Diehl:2011}.
In fact, the generation of low-entropy quantum states via engineered dissipation may play an essential role in the successful quantum simulation of, e.g., the ground state of the Fermi-Hubbard model away from half-filling, which would have far-reaching consequences for our understanding of high-temperature superconductivity.
Recently, the first small-scale quantum simulators for open quantum systems that employ tailored dissipation have been successfully realized in the laboratory \cite{Barreiro:2011,Schindler:2013}.

While the nonequilibrium stationary state in these systems is typically known by construction, far less is known about their real-time dynamics.
Accordingly, it is so far not possible to make clear statements on the efficiency and stability of dissipative state preparation protocols, i.e., on the time scales at which the nonequilibrium stationary state is established.
This is largely due to the fact that numerical methods based on the density matrix renormalization group (DMRG) \cite{Vidal:2004,White:2004,Verstraete:2004} and the quantum trajectory method \cite{Daley:2014fha} are limited to low-dimensional systems and short times.
Remarkably, there are rare instances in which the real-time dynamics of large open quantum systems can be solved exactly or semi-analytically \cite{Znidaric:2010,Horstmann:2013,Zunkovic:2014,Caspar:2015,Caspar:2016}.
For instance, the hierarchy of time evolution equations of correlation functions closes for a purely dissipative process that drives a system of hardcore bosons into a Bose-Einstein condensate. 
In this specific case, it was shown that the dissipative gap that determines the long-time behavior of the quantum many-body system shows an intriguing finite-size scaling as a function of dimensionality which results in a increase of efficiency in higher dimensions \cite{Caspar:2015,Caspar:2016}.
It is, however, unclear whether this strong dependence of the time scales on dimensionality is a generic feature of tailored dissipative processes or rather a special property of the considered state preparation protocol.

It remains a major challenge of contemporary theoretical physics to develop efficient methods to study the real-time dynamics of large quantum many-body systems in more than one spatial dimension.
In this work, we apply the phase space formulation of quantum dynamics using the Wigner-Weyl quantization to driven open quantum many-body systems.
This approach, which has been widely used for studying the real-time dynamics of closed quantum system (for reviews see, e.g., \cite{Blakie:2008,Polkovnikov:2009ys}), is particularly well-suited for bosonic systems for which the average occupation numbers is not too small.
In the simplest (semiclassical) truncated Wigner approximation (TWA), quantum fluctuations are taken into account in the initial state whereas the subsequent dynamics is governed by classical evolution equations.
This approximation is asymptotically exact at short times, however, the instant of time where it breaks down strongly depends on the details of the considered problem. 
In this work, we use the TWA to study how much of the dynamics of dissipative cooling is already captured by a semiclassical analysis.
This provides insight into the applicability of this method as well as into necessary extensions of the semiclassical analysis.

To this end, we derive the functional integral that governs the time evolution of the Wigner function, i.e., the representation of the density matrix in the coherent state phase space.
The leading order expansion of the Markovian dissipative action in the response field then defines the TWA, which is employed to study the growth of long-range correlation in real time. 
Unlike alternative techniques, the computational effort does not grow exponentially with the size of the Hilbert space so that numerical studies for large systems are feasible.
Specifically, we study a system of lattice bosons in two spatial dimensions that interacts with an engineered environment that drives a disordered, high-temperature initial state into a Bose-Einstein condensate \cite{Diehl:2008}.

The mechanism of dissipative cooling can be visualized as a process in which -- starting from a vortex condensate at initial times -- vortex-anitvortex pairs annihilate due to the interaction with the engineered environment.
The spontaneous formation of topological defects upon quenching a system at finite rate across a continuous phase transition lies at the heart of the Kibble-Zurek mechanism \cite{Kibble:1976,Zurek:1985}.
This effect has been experimentally tested on various platforms \cite{Bauerle:1996,Ruutu:1996,Carmi:2000,Monaco:2002,Maniv:2003} and correlations between the vortices has been studied \cite{Chu:2000xq,Rajarshi:2004,Golubchik:2010,Jelic:2011,Schole:2012}.
It is still an open question whether this mechanism also holds in the case of driven-dissipative systems, i.e., if the sweep through the critical point is generated by a dissipative coupling (dissipative Kibble-Zurek mechanism).

This paper is organized as follows:
In Sec.~\ref{sec:WignerWeyl} we review the phase space representation of quantum dynamics based on the Wigner-Weyl quantization.
We derive the Wigner-Weyl form of the Markovian dissipative action that governs the time evolution of the Wigner function.
Based on a leading order expansion in the response field, we determine the classical equations of motion and show how observables are calculated.
In Sec.~\ref{sec:DissCool} we discuss the dissipative state preparation protocol that drives a system of disordered lattice bosons into a Bose-Einstein condensate.
Using the TWA, we investigate the real-time dynamics of this process and we explain our findings in terms of the dynamics of vortex-antivortex pairs.
We observe that a finite vortex density survives the cooling process which, however, disagrees with the analytically constructed vortex-free Bose-Einstein condensate at asymptotic times.
This indicates that quantum fluctuations beyond the TWA need to be included to fully capture the physics of dissipative Bose-Einstein condensation.
Conclusions and an outlook are presented in Sec.~\ref{sec:Conclusions}.

\section{Wigner-Weyl quantization for open quantum systems}
\label{sec:WignerWeyl}

\subsection{Definitions}

In this section we briefly review basic definitions regarding coherent states that are needed afterwards, and we refer to \cite{Cahill:1969it,Hillery:1983ms} for further details.

We consider lattice bosons that obey the commutation relations $[a_i,a_j^{\dagger}]=\delta_{ij}$, with site labels $i,j\in\{1,\ldots,N\}$.
In the following, boldface symbols denote configurations on the spatial lattice $\bs\varphi=(\varphi_1,\ldots,\varphi_N)\in\mathbb{C}^N$ and the bar denotes complex conjugation $\overline{\bs{\varphi}}=(\varphi_1^*,\ldots,\varphi_N^*)$.
The inner product is written as $\overline{\bs{\eta}}\bs{\varphi}=\sum_{i=1}^{N}\eta_i^*\varphi_i$.

Coherent states $\ket{\bs\varphi}$ are defined as the eigenstates of the annihilation operators
\begin{equation}
 a_i\ket{\bs\varphi}=\varphi_i\ket{\bs\varphi} \ , 
\end{equation}
that can be expressed according to
\begin{equation}
 \ket{\bs{\varphi}}\equiv\exp(\bs{\varphi}\bs{a}^\dagger)\ket{\Omega}\ns{2}=\ns{2}\bigotimes_{i=1}^{N}\exp(\varphi_ia_i^\dagger)\ket{0}\ns{2}=\ns{2}\bigotimes_{i=1}^{N}\ket{\varphi_i} \ ,
\end{equation}
where $\ket{\Omega}=\bigotimes_{i=1}^{N}\ket{0}$ denotes the ground state in the multiparticle Fock space.
Coherent states are not orthogonal 
\begin{equation}
 \braket{\bs{\eta}}{\bs{\varphi}}=\ns{2}\prod_{i=1}^{N}\braket{\eta_i}{\varphi_i}=\ns{2}\prod_{i=1}^{N}e^{\eta_i^*\varphi_i}=e^{\overline{\bs{\eta}}\bs{\varphi}} \ ,
\end{equation}
and hence form an overcomplete basis. Consequently, the identity operator is given by
\begin{equation}
 \mathbbm{1}=\int{\ns{6}\mathcal{D}\bs{\varphi}\, e^{-|\boldsymbol{\varphi}|^2}\ketbra{\boldsymbol{\varphi}}{\boldsymbol{\varphi}}} \ ,
\end{equation}
where $\mathcal{D}\bs\varphi\equiv \pi^{-N}\prod_{i=1}^{N}(d\re\varphi_i)(d\im \varphi_i)$ denotes the integration measure .
As a consequence, the trace of any bosonic operator $O(\bs a, \bs a^\dagger)$ can be expressed in the phase space of coherent states according to
\begin{equation}
 \tr{O}=\int{\ns{6}\mathcal{D}\bs{\varphi}\, e^{-|\boldsymbol{\varphi}|^2}\bracket{\bs\varphi}{O}{\bs\varphi}} \ .
\end{equation}
The unitary displacement operator $D(\bs\varphi)$ is defined as
\begin{equation}
 \label{eq:disp}
 D(\bs\varphi)=D^\dagger(-\bs\varphi)=D^{-1}(-\bs\varphi)=\exp(\bs{\varphi}\bs a^\dagger-\overline{\bs{\varphi}}\bs{a}) \ ,
\end{equation}
such that $D(\bs\varphi)\ket{\Omega}=e^{-\frac{1}{2}|\bs\varphi|^2}\ket{\bs\varphi}$. 
Displacement operators with different arguments fulfill the composition rule
\begin{equation}
 \label{eq:comp}
 D(\bs\varphi)D(\bs\eta)=D(\bs\varphi+\bs\eta)e^{\frac{1}{2}\left(\overline{\bs\eta}\bs\varphi-\overline{\bs\varphi}\bs\eta\right)} 
\end{equation}
and are orthogonal in the sense that
\begin{equation}
 \label{eq:orth}
 \tr{D(\bs\varphi)D^\dagger(\bs\eta)}=\pi^N \delta(\bs\varphi-\bs\eta) \ ,
\end{equation}
where the complex Dirac $\delta$-function is defined according to
\begin{align}
 \label{eq:delta}
 \delta(\bs\varphi)&=\prod_{i=1}^{N}\delta(\re\varphi_i)\delta(\im\varphi_i)=\frac{1}{\pi^N}\int{\ns{6}\mathcal{D}\bs\eta\, e^{-\overline{\bs\varphi}\bs{\eta}+\overline{\bs{\eta}}\bs\varphi}} \ .
\end{align}
Most important, any bounded operator $O(\bs a,\bs a^\dagger)$ with finite Hilbert-Schmidt norm, $||O||^2=\tr{O^\dagger O}<\infty$, can be expanded in terms of displacement operators
\begin{equation}
 \label{eq:op}
 F\ns{5}=\ns{5}\int{\ns{6}\mathcal{D}\bs\eta\tr{D(\bs\eta)O}D^{\dagger}(\bs\eta)}\ns{5}=\ns{5}\int{\ns{6}\mathcal{D}\bs\eta\, \chi_O(\bs\eta) D^{\dagger}(\bs\eta)} \ .
\end{equation}
The quantity $\chi_O(\bs\eta)=\tr{D(\bs\eta)O}$ is called the characteristic function of the operator $O(\bs a, \bs a^\dagger)$.
The Fourier transform of the characteristic function defines its Weyl symbol:
\begin{subequations}
\begin{align}
 \label{eq:weyl}
 O_W(\bs\varphi)&=\int{\ns{6}\mathcal{D}\bs\eta\,e^{-\overline{\bs\varphi}\bs\eta+\overline{\bs\eta}\bs\varphi}\chi_O(\bs\eta)} \ , \\
 \chi_O(\bs\eta)&=\int{\ns{6}\mathcal{D}\bs\varphi\,e^{-\overline{\bs\eta}\bs\varphi+\overline{\bs\varphi}\bs\eta} O_W(\bs\varphi)} \ .
 \label{eq:char}
\end{align}
\end{subequations}
The Weyl symbol of the density matrix $\rho(t)$ is known as the Wigner function $\rho_W(\bs\varphi,t)$.
Most important, expectation values of bosonic operators $O(\bs a, \bs a^\dagger)$ are calculated in the phase space of coherent states according to
\begin{equation}
 \label{eq:expval}
 \langle O \rangle(t) = \tr{\rho(t)O}=\int{\ns{6}\mathcal{D}\bs{\varphi}\,\rho_W(\bs\varphi,t)O_W(\bs\varphi)} \ .
\end{equation}

\subsection{Markovian quantum master equation}

We consider a system whose evolution of the density matrix $\rho(t)$ is governed by a Markovian quantum master equation in Lindblad form \cite{Gorini:1976,Lindblad:1975,Breuer:2007}
\begin{equation}
 \label{eq:lindblad}
 \frac{d}{dt}\rho(t)\ns{4}=\ns{4}-i[H,\rho(t)]+\sum_{\alpha}\ns{4}\gamma_\alpha\Big(\ns{4} L_{\alpha}\rho(t) L_{\alpha}^\dagger-\frac{1}{2}\{L_{\alpha}^\dagger L_{\alpha},\rho(t)\}\ns{4}\Big) \ ,
\end{equation}
where $H$ is the system Hamiltonian, and $L_{\alpha}$ are a set of quantum jump operators, which describe the interaction of the bosons with the environment and act at a rate $\gamma_\alpha$.
This is the most general non-unitary, time-local, Markovian evolution equation that preserves the basic properties of Hermiticity and positive semi-definiteness of the density matrix.
The time evolution equation for the Wigner function $\rho_W(\bs\varphi,t)$ is obtained by calculating the Wigner transform of \eqref{eq:lindblad}, which requires to determine the Weyl symbols of products of two operators (Hamiltonian terms) and three operators (dissipative terms).
\vspace{-0.2cm}
\subsubsection{Product formula for Weyl symbols}

To calculate the Weyl symbol of the product of two operators $(FG)_W(\bs\varphi)$, we first take its definition \eqref{eq:weyl} and then use the operator expansion \eqref{eq:op} twice.
The resulting trace over displacement operators can be calculated by taking into account the composition rule \eqref{eq:comp} along with the orthogonality relation \eqref{eq:orth}:
\begin{align}
 \tr{D^\dagger(\bs\lambda)D^\dagger(\bs\xi)D(\bs\eta)}\ns{3}=\ns{3}\pi^Ne^{\frac{1}{2}(\overline{\bs\xi}\bs\lambda-\overline{\bs\lambda}\bs\xi)}\delta(\bs\eta\ns{3}-\ns{3}\bs\lambda\ns{3}-\ns{3}\bs\xi) \, .
\end{align}
By integrating out the variable $\bs\eta$ we obtain:
\begin{align}
 &\ns{6}(FG)_W(\bs\varphi)\ns{4}=\ns{4}\int{\ns{6}\mathcal{D}\bs\eta\,e^{-\overline{\bs\varphi}\bs\eta+\overline{\bs\eta}\bs\varphi}\chi_{FG}(\bs\eta)}= \notag \\
 &\ns{8}=\ns{5}\int{\ns{6}\mathcal{D}\bs\lambda\mathcal{D}\bs\xi\,e^{-\overline{\bs\varphi}(\bs\lambda+\bs\xi)+(\overline{\bs\lambda}+\overline{\bs\xi})\bs\varphi}e^{\frac{1}{2}(\overline{\bs\xi}\bs\lambda-\overline{\bs\lambda}\bs\xi)}\chi_{F}(\bs\lambda)\chi_{G}(\bs\xi)} \, .
\end{align}
If we then again use the Fourier representation of the characteristic functions \eqref{eq:char} along with the integral representation of the complex Dirac $\delta$-function \eqref{eq:delta}, we finally obtain two equivalent integral expressions
\begin{subequations}
\label{eq:twoop}
\begin{align}
 &(FG)_W(\bs\varphi)= \notag \\
 &=\ns{5}\int{\ns{6}\mathcal{D}\bs\lambda\mathcal{D}\bs\eta\,e^{-\overline{\bs\eta}(\bs\varphi-\bs\lambda)+(\overline{\bs\varphi}-\overline{\bs\lambda})\bs\eta}F_W(\bs\varphi+\tfrac{1}{2}\bs\eta)G_W(\bs\lambda)} \\
 &=\ns{5}\int{\ns{6}\mathcal{D}\bs\lambda\mathcal{D}\bs\eta\,e^{-\overline{\bs\eta}(\bs\varphi-\bs\lambda)+(\overline{\bs\varphi}-\overline{\bs\lambda})\bs\eta}F_W(\bs\lambda)G_W(\bs\varphi-\tfrac{1}{2}\bs\eta)}
\end{align}
\end{subequations}
In an analogous way we calculate the Weyl symbol of the product of three operators $(FGH)_W(\bs\varphi)$.
A lengthy but straightforward calculation yields three equivalent integral expressions
\begin{widetext}
\begin{subequations}
\label{eq:threeop}
\begin{align}
 (FGH)_W(\bs\varphi) &=\int{\ns{6}\mathcal{D}\bs\lambda\mathcal{D}\bs\eta\,e^{-\overline{\bs\eta}(\bs\varphi-\bs\lambda)+(\overline{\bs\varphi}-\overline{\bs\lambda})\bs\eta}F_W(\bs\varphi+\tfrac{1}{2}\bs\eta)G_W(\bs\lambda+\tfrac{1}{2}\bs\eta)H_W(\bs\lambda)}  \\
                     &=\int{\ns{6}\mathcal{D}\bs\lambda\mathcal{D}\bs\eta\,e^{-\overline{\bs\eta}(\bs\varphi-\bs\lambda)+(\overline{\bs\varphi}-\overline{\bs\lambda})\bs\eta}F_W(\tfrac{1}{2}\bs\varphi+\tfrac{1}{2}\bs\lambda+\tfrac{1}{2}\bs\eta)G_W(\bs\lambda)H_W(\tfrac{1}{2}\bs\varphi+\tfrac{1}{2}\bs\lambda-\tfrac{1}{2}\bs\eta)}  \\ 
                     &=\int{\ns{6}\mathcal{D}\bs\lambda\mathcal{D}\bs\eta\,e^{-\overline{\bs\eta}(\bs\varphi-\bs\lambda)+(\overline{\bs\varphi}-\overline{\bs\lambda})\bs\eta}F_W(\bs\lambda)G_W(\bs\lambda-\tfrac{1}{2}\bs\eta)H_W(\bs\varphi-\tfrac{1}{2}\bs\eta)} 
\end{align}
\end{subequations}
\end{widetext}

\subsubsection{Time-discretized evolution}
We consider a time discretization of the master equation \eqref{eq:lindblad}.
For $\epsilon=t_1-t_0\to0$, it is given by
\begin{align}
 \label{eq:lindblad2}
 \rho(t_1)&=\rho(t_0)-i\epsilon H\rho(t_0)+i\epsilon \rho(t_0)H\notag \\
         &+\ns{2}\epsilon \sum_{\alpha}\ns{4}\gamma_\alpha\Big(\ns{2}L_\alpha\rho(t_0) L_\alpha^\dagger\ns{1}-\ns{1}\frac{1}{2}\{L_{\alpha}^\dagger L_\alpha,\rho(t_0)\}\Big) \, .
\end{align} 
This operator equation can be directly translated into an equation for its characteristic functions or, equivalently, its Weyl symbols. 
The Weyl symbol of its left-hand side gives the time-evolved Wigner function $\rho_W(\bs\varphi_1,t_1)$.
The first term on the right-hand side is given by the Wigner function $\rho_W(\bs\varphi_1,t_0)$.
Moreover, we employ the product formula for Weyl symbols \eqref{eq:twoop} and \eqref{eq:threeop} to calculate the remaining terms on the right-hand side.

\begin{widetext}
\noindent First, we note that the linear term can be rewritten with the aid of the complex Dirac $\delta$-function \eqref{eq:delta} according to:
\begin{align}
 \rho_W(\bs\varphi_1,t_0)=\ns{5}\int{\ns{6}\mathcal{D}\bs\varphi_0}\,\delta(\bs\varphi_0-\bs\varphi_1)\rho_W(\bs\varphi_0,t_0)=\ns{5}\int{\ns{6}\mathcal{D}\bs\varphi_0}\mathcal{D}\bs\eta_1\,e^{-\overline{\bs{\eta}}_1(\bs\varphi_1-\bs\varphi_0)+(\overline{\bs\varphi}_1-\overline{\bs\varphi}_0)\bs{\eta_1}}\rho_W(\bs\varphi_0,t_0) \, .
\end{align}
We further note that the exponential factors in \eqref{eq:twoop} and \eqref{eq:threeop} are identical.
It is hence possible to write the Weyl symbols on the right-hand side of \eqref{eq:lindblad2} in a unified manner that comprises the Wigner function at the preceding time step $\rho_W(\bs\varphi_0,t_0)$. 
Specifically, we obtain for the Wigner function after an infinitesimal time step:
\begin{align}
 \rho_W(\bs\varphi_1,t_1)&=\ns{5}\int{\ns{6}\mathcal{D}\bs\varphi_0}\mathcal{D}\bs\eta_1\,e^{-\overline{\bs{\eta}}_1\ns{2}(\bs\varphi_1-\bs\varphi_0)+(\overline{\bs\varphi}_1-\overline{\bs\varphi}_0)\bs{\eta_1}}\Big[1-i\epsilon \mathcal{H}(\bs\varphi_1,\bs\eta_1)+\epsilon\sum_{\alpha}\gamma_\alpha\mathcal{L}_\alpha(\bs\varphi_0,\bs\varphi_1,\bs\eta_1)\Big]\rho_W(\bs\varphi_0,t_0) \notag \\
                         &=\ns{5}\int{\ns{6}\mathcal{D}\bs\varphi_0}\mathcal{D}\bs\eta_1\,e^{-\overline{\bs{\eta}}_1\ns{2}(\bs\varphi_1-\bs\varphi_0)+(\overline{\bs\varphi}_1-\overline{\bs\varphi}_0)\bs{\eta_1}}\exp\Big[-i\epsilon \mathcal{H}(\bs\varphi_1,\bs\eta_1)+\epsilon\sum_{\alpha}\gamma_\alpha\mathcal{L}_\alpha(\bs\varphi_0,\bs\varphi_1,\bs\eta_1)\Big]\rho_W(\bs\varphi_0,t_0) \ .
\end{align}
In the second line we exponentiated the expression in the bracket, which is exact to linear order in $\epsilon$.
This procedure can be reiterated from initial time $t_0$ to final time $t_M$, where we assume $\epsilon=(t_M-t_0)/M\to0$ as $M\to\infty$:
\begin{align}
 \label{eq:wig_evolved}
 \rho_W(\bs\varphi_M,t_M)=\ns{5}\int{\ns{6}\prod_{k=1}^{M}\mathcal{D}\bs\varphi_{k-1}\mathcal{D}\bs\eta_{k}\,e^{-\overline{\bs{\eta}}_k\ns{2}(\bs\varphi_k-\bs\varphi_{k-1})+(\overline{\bs\varphi}_k-\overline{\bs\varphi}_{k-1})\bs{\eta}_k}\exp\Big[\ns{4}-\ns{4}i\epsilon \mathcal{H}(\bs\varphi_k,\bs\eta_k)\ns{4}+\ns{4}\epsilon\sum_{\alpha}\gamma_\alpha\mathcal{L}_\alpha(\bs\varphi_{k-1},\bs\varphi_k,\bs\eta_k)\Big]\rho_W(\bs\varphi_0,t_0)}
\end{align}
The variables $\bs\varphi_k$ and $\bs\eta_k$ have a natural interpretation as classical and response field, respectively, on the Schwinger-Keldysh closed time contour \cite{Schwinger:1960qe,Keldysh:1964ud,Kamenev:2011}.
Moreover, we introduced the Hamiltonian and dissipative contributions
\begin{subequations}
\label{eq:hamdiss}
\begin{align}
 \mathcal{H}(\bs\varphi_k,\bs\eta_k)&=H_W(\bs\varphi_k^+)-H_W(\bs\varphi_k^-) \, , \\
 \mathcal{L}_\alpha(\bs\varphi_{k-1},\bs\varphi_k,\bs\eta_k)&=L_{\alpha,W}(\bs\varphi_{k}^+ \ns{3}-\ns{2} \tfrac{1}{2}\epsilon\dot{\bs\varphi}_{k})L^*_{\alpha,W}(\bs\varphi_{k}^- \ns{3}-\ns{2} \tfrac{1}{2}\epsilon\dot{\bs\varphi}_{k})\ns{3}-\ns{3}\frac{1}{2}\big[L^*_{\alpha,W}(\bs\varphi_k^+)L_{\alpha,W}(\bs\varphi_k^+ \ns{3}-\ns{2} \epsilon\dot{\bs\varphi}_{k})\ns{3}+\ns{3}L^*_{\alpha,W}(\bs\varphi_k^- \ns{3}-\ns{2} \epsilon\dot{\bs\varphi}_k)L_{\alpha,W}(\bs\varphi_k^-)\big] \notag \\
 &=L_{\alpha,W}(\bs\varphi_{k}^+ \ns{3}-\ns{2} \tfrac{1}{2}\epsilon\dot{\bs\varphi}_{k})L^*_{\alpha,W}(\bs\varphi_{k}^- \ns{3}-\ns{2} \tfrac{1}{2}\epsilon\dot{\bs\varphi}_{k})\ns{3}-\ns{3}\frac{1}{2}\big[(L^\dagger_{\alpha}L_\alpha)_W(\bs\varphi_k^+)+(L^\dagger_{\alpha}L_{\alpha})_W(\bs\varphi_k^-)\big] \ .
\end{align}
\end{subequations}
\end{widetext}
The field variables on the upper $(+)$ and lower $(-)$ time branch are defined as
\begin{equation}
 \bs\varphi_k^\pm=\bs\varphi_k\pm\frac{1}{2}\bs\eta_k \ ,
\end{equation}
and for later convenience we introduced 
\begin{subequations}
 \begin{align}
  \widetilde{\bs\varphi}_{k}&=\frac{\bs\varphi_{k}+\bs\varphi_{k-1}}{2} \ , \\
  \dot{\bs\varphi}_{k}&=\frac{\bs\varphi_{k}-\bs\varphi_{k-1}}{\epsilon} \ . 
 \end{align}
\end{subequations}
The Hamiltonian terms appear as the difference between contributions from the upper and the lower time branches
\begin{equation}
 [H,\rho(t_{k-1})] \longrightarrow H_W(\bs\varphi_k^+)-H_W(\bs\varphi_k^-) \ .
\end{equation}
The anticommutator term also factorizes into the sum of two contributions on the upper and the lower time branches
\begin{equation}
 \{L_{\alpha}^\dagger L_\alpha,\rho(t_{k-1})\}\longrightarrow
 (L_\alpha^\dagger L_\alpha)_W(\bs\varphi_k^+)+(L_\alpha^\dagger L_\alpha)_W(\bs\varphi_k^-) \ .
\end{equation}
The so-called recycling term, however, does not show this factorization but rather correlates the upper and the lower time branches
\begin{equation}
 L_\alpha\rho(t_{k-1})  L_\alpha^\dagger \longrightarrow L_{\alpha,W}(\bs\varphi_{k}^+ - \tfrac{1}{2}\epsilon\dot{\bs\varphi}_{k})L^*_{\alpha,W}(\bs\varphi_{k}^- - \tfrac{1}{2}\epsilon\dot{\bs\varphi}_{k}) \ .
\end{equation}
Loosely speaking, an operator that acts from the left on the density matrix $\rho(t)$ appears as its corresponding Weyl symbol on the upper time branch $(+)$, whereas an operator that act from the right on the density matrix $\rho(t)$ appear as its corresponding Weyl symbol on the lower time branch $(-)$.
The terms proportional to $\epsilon\dot{\bs\varphi}_k$ can be regarded as a time regularization in the continuum limit $\epsilon\to0$.
We note that this ordering closely resembles the regularization of the Markovian dissipative action in the Schwinger-Keldysh formalism \cite{Sieberer:2013lwa,Maghrebi:2016}.
\vspace{-0.4cm}
\subsubsection{Truncated Wigner Approximation (TWA)}\label{sec:twa}
\vspace{-0.3cm}
In general, it is not possible to determine the time-evolved Wigner function \eqref{eq:wig_evolved} analytically.
A possible approximation, however, is found by expanding the Hamiltonian and dissipative terms in \eqref{eq:hamdiss} up to linear order in the response variables $\bs\eta_{k}=(\eta_{k,1},\ldots,\eta_{k,N})$.
The Hamiltonian terms yield
\begin{align}
 \mathcal{H}(\bs\varphi_k,\bs\eta_k)=\frac{\partial H_W(\bs\varphi_k)}{\partial \varphi_{k,i}}\eta_{k,i}+\frac{\partial H_W(\bs\varphi_k)}{\partial \varphi_{k,i}^*}\eta_{k,i}^*+\mathcal{O}(\bs\eta^3) \ ,
\end{align}
where the summation over the spatial indices $i$ is implied.
\begin{widetext}
\noindent On the other hand, the dissipative contributions give
\begin{align}
 &\mathcal{L}_\alpha(\bs\varphi_{k-1},\bs\varphi_k,\bs\eta_k)=\mathcal{L}_{\alpha,0}(\bs\varphi_{k-1},\bs\varphi_{k})+\mathcal{L}_{\alpha,1,i}(\bs\varphi_{k-1},\bs\varphi_{k})\eta_{k,i}-\mathcal{L}^*_{\alpha,1,i}(\bs\varphi_{k-1},\bs\varphi_{k})\eta^*_{k,i}+\mathcal{O}(\bs\eta^2) \ ,
\end{align}
with
\begin{subequations}
\begin{align}
 \mathcal{L}_{\alpha,0}(\bs\varphi_{k-1},\bs\varphi_k)&=L^*_{\alpha,W}(\widetilde{\bs\varphi}_k)L_{\alpha,W}(\widetilde{\bs\varphi}_k)-\frac{1}{2}\big[L^*_{\alpha,W}(\bs\varphi_k)L_{\alpha,W}(\bs\varphi_{k-1})+L^*_{\alpha,W}(\bs\varphi_{k-1})L_{\alpha,W}(\bs\varphi_k)\big] \ , \\
 \mathcal{L}_{\alpha,1,i}(\bs\varphi_{k-1},\bs\varphi_k)&=\frac{1}{2}\Big[L^*_{\alpha,W}(\widetilde{\bs\varphi}_k)\frac{\partial L_{\alpha,W}(\widetilde{\bs\varphi}_k)}{\partial\widetilde{\varphi}_{k,i}}-L_{\alpha,W}(\widetilde{\bs\varphi}_k)\frac{\partial L^*_{\alpha,W}(\widetilde{\bs\varphi}_k)}{\partial\widetilde{\varphi}_{k,i}}\Big]+\frac{1}{4}\Big[\ns{2}L^*_{\alpha,W}(\bs\varphi_{k-1})\frac{\partial L_{\alpha,W}(\bs\varphi_k)}{\partial\varphi_{k,i}} \notag \\
 &-L_{\alpha,W}(\bs\varphi_{k-1})\frac{\partial L^*_{\alpha,W}(\bs\varphi_k)}{\partial\varphi_{k,i}}-L^*_{\alpha,W}(\bs\varphi_{k})\frac{\partial L_{\alpha,W}(\bs\varphi_{k-1})}{\partial\varphi_{k-1,i}}+L_{\alpha,W}(\bs\varphi_{k})\frac{\partial L^*_{\alpha,W}(\bs\varphi_{k-1})}{\partial\varphi_{k-1,i}}\Big] \, .
\end{align}
\end{subequations}
Within this approximation, the response fields $\bs\eta_k$ appear only linearly in the expression for the time-evolved Wigner function \eqref{eq:wig_evolved}.
Accordingly, they can be integrated out exactly in terms of a product of complex Dirac $\delta$-functions \eqref{eq:delta}
\begin{align}
 \label{eq:wig_twa1}
 \rho_W(\bs\varphi_M,t_M)&=\int{\ns{3}\prod_{k=1}^{M}\widetilde{\mathcal{D}}\bs\varphi_{k-1}\,\rho_W(\bs\varphi_0,t_0)\exp\Big[\epsilon\sum_{\alpha}\gamma_\alpha\mathcal{L}_{\alpha,0}(\bs\varphi_{k-1},\bs\varphi_{k})\Big]\delta(\bs{\mathcal{F}}(\bs\varphi_{k-1},\bs\varphi_{k}))} \ ,
\end{align}
with the modified integration measure $\widetilde{\mathcal{D}}\bs\varphi_{k}=\prod_{i=1}^{N}(d\re\varphi_{k,i})(d\im \varphi_{k,i})$.
The argument of the complex Dirac $\delta$-functions encode the classical equations of motion 
\begin{equation}
 \label{eq:eom_disc}
 \mathcal{F}_i(\bs\varphi_{k-1},\bs\varphi_{k})=\varphi_{k-1,i}-\varphi_{k,i}-i\epsilon\frac{\partial H_W(\bs\varphi_{k})}{\partial\varphi_{k,i}^*}-\epsilon\sum_{\alpha}\gamma_\alpha\mathcal{L}^*_{\alpha,1,i}(\bs\varphi_{k-1},\bs\varphi_{k})\stackrel{!}{=}0 \ .
\end{equation}
In this form, they are implicit finite difference equations for $\bs\varphi_{k}$ in terms of the previous value $\bs\varphi_{k-1}$, such that $\bs\varphi_{k}=\bs{\mathcal{G}}(\bs\varphi_{k-1})$.
For actual numerical simulations, it will be helpful to take the time-continuum limit $\epsilon\to0$.\\
\end{widetext} 
                                                                                                                                                                                        
\subsubsection{Time-continuum limit}

In the time-continuum limit, the classical equation of motion \eqref{eq:eom_disc} determines the time derivatives of the field variables
\begin{align}
 \dot{\varphi}_{k,i}&=\lim_{\epsilon\to0}\frac{\varphi_{k,i}-\varphi_{k-1,i}}{\epsilon} \notag \\
 &=-i\frac{\partial H_W(\bs\varphi_{k})}{\partial\varphi_{k,i}^*}-\sum_{\alpha}\gamma_\alpha\mathcal{L}^*_{\alpha,1,i}(\bs\varphi_{k-1},\bs\varphi_{k}) \ .
\end{align}
In the following we assume that the field configurations $\bs\varphi_k$ are sufficiently smooth so that the time derivative remains finite at any instant of time, $|\dot{\bs\varphi}_{k}|<\infty$.
Under this assumption, we find that $\mathcal{L}_{\alpha,0}(\bs\varphi_{k-1},\bs\varphi_{k})$ vanishes quadratically for $\epsilon\to0$ by expanding it around $\widetilde{\bs\varphi}_{k}$:
\begin{equation}
 \mathcal{L}_{\alpha,0}(\bs\varphi_{k-1},\bs\varphi_{k})=\mathcal{O}(\epsilon^2\dot{\bs\varphi}^2_{k}) \ .
\end{equation}
Accordingly, the exponential factor in \eqref{eq:wig_twa1} can be disregarded in the time-continuum limit
\begin{equation}
 \exp\Big[\epsilon\sum_{\alpha}\gamma_\alpha\mathcal{L}_{\alpha,0}(\bs\varphi_{k-1},\bs\varphi_{k})\Big]\stackrel{\epsilon\to0}{\Longrightarrow} 1 \ .
\end{equation}
Moreover, in order to integrate over the remaining variables in \eqref{eq:wig_twa1}, we have to account for an inverse Jacobian 
\begin{align}
 \delta(\bs{\mathcal{F}}(\bs\varphi_{k-1},\bs\varphi_{k}))&=
 \left[\det \mathcal{J}_k\right]^{-1}\delta(\bs\varphi_{k}-\bs{\mathcal{G}}(\bs\varphi_{k-1})) \ ,
\end{align}
where we assumed that the implicit evolution equation, $\bs{\mathcal{F}}(\bs\varphi_{k-1},\bs\varphi_{k})=0$, has a unique solution, $\bs\varphi_{k}=\bs{\mathcal{G}}(\bs\varphi_{k-1})$.
In fact, the Jacobi matrix takes the form
\begin{equation}
 \mathcal{J}_k=\frac{\partial(\re \bs{\mathcal{F}},\im \bs{\mathcal{F}})}{\partial(\re\bs\varphi_{k},\im\bs\varphi_k)}=\frac{\partial(\bs{\mathcal{F}},\bs{\mathcal{F}}^*)}{\partial(\bs\varphi_{k},\bs\varphi_k^*)}=-\mathbbm{1}-\epsilon\mathcal{X}_k \ .
\end{equation}
In the time-continuum limit $\epsilon\to0$, we expand the determinant according to
\begin{equation}
 \det \mathcal{J}_k = \det(\mathbbm{1}+\epsilon\mathcal{X}_k)=1+\epsilon\tr{\mathcal{X}_k}+\mathcal{O}(\epsilon^2) \ .
\end{equation}
The Hamiltonian contributions exactly cancel whereas the dissipative terms account for a contribution
\begin{align}
 &\operatorname{Tr}[\mathcal{X}_k]=2\sum_{i=1}^{N}\sum_{\alpha}\gamma_{\alpha}\ns{4}\re\ns{4}\Big[\frac{\partial \mathcal{L}_{\alpha,1,i}(\bs\varphi_{k-1},\bs\varphi_k)}{\partial\varphi^*_{k,i}}\Big]=\mathcal{O}(\epsilon\dot{\bs\varphi}_k) \ ,
\end{align}
where we expanded again all terms around $\widetilde{\bs\varphi}_{k}$.
Due to the fact that the nontrivial terms vanish quadratically for $\epsilon\to0$, the inverse Jacobian does not contribute in the time-continuum limit
\begin{equation}
 \left[\det \mathcal{J}_k\right]^{-1}=\left[1+\mathcal{O}(\epsilon^2\dot{\bs\varphi}_k)\right]^{-1}\stackrel{\epsilon\to0}{\Longrightarrow} 1 \, .
\end{equation}
Based on \eqref{eq:expval}, the expectation value of a bosonic operator $O(\bs a, \bs a^\dagger)$ in the TWA for $\epsilon\to0$ is given by
\begin{align}
 \label{eq:obs}
 \langle O \rangle(t_M)&=\int{\mathcal{D}\bs\varphi_{0}\,\rho_W(\bs\varphi_0,t_0)} \notag \\
 &\times\int{\prod_{k=1}^{M}\widetilde{\mathcal{D}}\bs\varphi_{k}\delta(\bs\varphi_{k+1}-\bs{\mathcal{G}}(\bs\varphi_{k}))O_W(\bs\varphi_M)} \, .
\end{align}
As for closed quantum systems \cite{Polkovnikov:2009ys,Berges:2007ym}, this can be efficiently simulated by sampling configurations from the initial Wigner function $\rho_W(\bs{\varphi}_0,t_0)$, evolving each ensemble member according to the classical equation of motion \eqref{eq:eom_disc}, and then taking the average over the individual contributions.
By introducing the continuous time notation $\bs\varphi(t)\equiv\widetilde{\bs\varphi}_k$, the discrete expression \eqref{eq:obs} is conveniently written as
\begin{equation}
 \label{eq:obs2}
 \langle O\rangle(t)=\ns{4}\int{\ns{4}\mathcal{D}\bs{\varphi}_0}\rho_W(\bs{\varphi}_0,t_0)\ns{24}\int_{\bs\varphi(t_0)=\bs\varphi_0}{\ns{24}\mathcal{D}\bs{\varphi}(t)\delta(e.o.m.) O_W(\bs\varphi)}\, ,
\end{equation}
where $\delta(e.o.m.)$ indicates that the individual field configurations $\bs\varphi(t)$ obey the classical equation of motion in the time-continuum limit
\begin{align}
 \label{eq:eom}
 i\dot{\varphi}_i(t)&=\frac{\partial H_W(\bs\varphi)}{\partial \varphi^*_i}+\frac{i}{2}\sum_{\alpha}\gamma_\alpha \notag \\
 &\times\Big(L^*_{\alpha,W}(\bs\varphi)\frac{L_{\alpha,W}(\bs\varphi)}{\partial \varphi^*_i}-L_{\alpha,W}(\bs\varphi)\frac{L^*_{\alpha,W}(\bs\varphi)}{\partial \varphi^*_i}\Big) \ .
\end{align}
We emphasize that the dissipative contributions are only nonvanishing for non-Hermitean quantum jump operators $L_{\alpha}\neq L_{\alpha}^\dagger$, such that $L_{\alpha,W}(\bs\varphi)\neq L_{\alpha,W}^*(\bs\varphi)$.

\section{Dissipative Bose-Einstein condensation} 
\label{sec:DissCool}

\subsection{Truncated Wigner Approximation (TWA)}

Quantum many-body systems of bosons can be driven into a Bose-Einstein condensate by engineering the coupling between the quantum system and its environment \cite{Diehl:2008}.
This mechanism has been experimentally realized on small systems for a discretized time evolution using stroboscopic dynamics via dynamical maps \cite{Schindler:2013}.
The quantum jump operators
\begin{equation}
 \label{eq:jump}
 L_{ij}=(a_i^\dagger+a_j^\dagger)(a_i-a_j) 
\end{equation}
that are defined on nearest-neighbor lattice sites $\langle i,j\rangle$ annihilate antisymmetric, out-of-phase superpositions of bosons and turn them into symmetric, in-phase superpositions.
For jump operators that act isotropically on each adjacent pair $\langle i,j\rangle$, the local locking of the bosonic phases results in a global phase coherence, corresponding to a Bose-Einstein condensate.

To study the real-time dynamics of this process for large systems, we employ the TWA for open quantum systems as derived in the previous section.
In this work we are primarily interested in the dynamics due to the dissipative coupling.
Accordingly, we restrict ourselves to a purely dissipative system and set $H=0$.
We emphasize, however, that it is straightforward to include Hamiltonian contributions to the dynamics.

To calculate the Weyl symbol of the jump operator \eqref{eq:jump}, we first note the displacement operator $D(\bs\eta)$ can be written with the aid of the Baker-Campbell-Hausdorff formula as
\begin{align}
 D(\bs\eta)&=e^{\frac{1}{2}|\bs\eta|^2}\exp(-\overline{\bs\eta}\bs{a})\exp(\bs\eta \bs{a}^\dagger) \notag \\
 &=e^{-\frac{1}{2}|\bs\eta|^2}\exp(\bs\eta \bs{a}^\dagger)\exp(-\overline{\bs\eta}\bs{a})  \ .
\end{align}
Consequently, the action of a creation or annihilation operator on the displacement operator $D(\bs\eta)$ is given by
\begin{subequations} 
\begin{align}
  a_i^\dagger D(\bs\eta)&=\Big(\frac{1}{2}\eta_i^*+\frac{\partial}{\partial \eta_i}\Big)D(\bs\eta) \ , \\
  a_i D(\bs\eta)&=\Big(\frac{1}{2}\eta_i-\frac{\partial}{\partial \eta_i^*}\Big)D(\bs\eta) \ ,
\end{align}
\end{subequations}
and we also obtain
\begin{align}
 &a_i^\dagger a_j D(\bs\eta)=\Big(\frac{1}{2}\eta_j-\frac{\partial}{\partial \eta_j^*}\Big)\Big(\frac{1}{2}\eta_i^*+\frac{\partial}{\partial \eta_i}\Big)D(\bs\eta) \notag  \\
 &=\Big(\frac{1}{4}\eta_i^*\eta_j\ns{2}+\ns{2}\frac{1}{2}\eta_j\frac{\partial}{\partial \eta_i}\ns{2}-\ns{2}\frac{1}{2}\eta_i^*\frac{\partial}{\partial \eta_j^*}\ns{2}-\ns{2}\frac{\partial^2}{\partial\eta_j^*\partial\eta_i}\ns{2}-\ns{2}\frac{1}{2}\delta_{ij}\Big)D(\bs\eta) \, .
\end{align}
Accordingly, the Weyl symbol of the operator $a_i^\dagger a_j$ reads
\begin{align}
 &(a_i^\dagger a_j)_W(\bs\varphi)=\int{\mathcal{D}\bs\eta e^{-\overline{\bs\varphi}\bs\eta+\overline{\bs\eta}\bs\varphi}\tr{a_i^\dagger a_j D(\bs\eta)}} \notag \\
 &=\pi^N\ns{8}\int{\ns{6}\mathcal{D}\bs\eta e^{-\overline{\bs\varphi}\bs\eta+\overline{\bs\eta}\bs\varphi}\Big(\frac{1}{2}\eta_j\frac{\partial}{\partial \eta_i}\ns{2}-\ns{2}\frac{1}{2}\eta_i^*\frac{\partial}{\partial \eta_j^*}\ns{2}-\ns{2}\frac{\partial^2}{\partial\eta_j^*\partial\eta_i}} \notag \\
 &\quad+\frac{1}{4}\eta_i^*\eta_j\ns{2}-\ns{2}\frac{1}{2}\delta_{ij}\Big)\delta(\bs\eta)=\varphi_i^*\varphi_j-\frac{1}{2}\delta_{ij} \ .
\end{align}
The Weyl symbol of the jump operator \eqref{eq:jump} is given by
\begin{equation}
 L_{ij,W}(\boldsymbol{\varphi})=|\varphi_i|^2-|\varphi_j|^2+\varphi_j^*\varphi_i-\varphi_i^*\varphi_j \ ,
\end{equation}
so that the classical equation of motion \eqref{eq:eom} reads
\begin{equation}
 \label{eq:eom_cooling}
 \dot{\varphi}_j(t)=-\gamma \sum_{i|\langle i,j\rangle }\varphi_{i}^*(t)\left[\varphi_{j}^2(t)-\varphi_{i}^2(t)\right] \ .
\end{equation}
The sum extends over all lattices sites $i$ that are adjacent to the lattice site $j$.
Moreover, we assumed that the interaction rate is independent of the specific pair of sites, $\gamma_{ij}=\gamma$.
We note that the lattice equation of motion \eqref{eq:eom_cooling} can be regarded as the leading term of the nonlinear partial differential equation
\begin{equation}
 \label{eq:eom_cont}
 \dot{\varphi}(\bs x,t)=\widetilde{\gamma}\, [\varphi^*(\bs x,t)\Delta\varphi^2(\bs x,t)+4\varphi(\bs x,t)|\nabla \varphi(\bs x,t)|^2 ] \, ,
\end{equation}
with $\widetilde{\gamma}=a^2\gamma$ and the continuous-space field variable $\varphi(\bs x,t)\equiv\varphi_{i}(t)$.
\newpage
We emphasize that the homogeneous, phase-coherent field configuration 
\begin{equation}
 \lim_{t\to\infty}\varphi_{j}(t)=\Phi_0 e^{i\Theta_0}
\end{equation}
with $\Theta_0\in[0,2\pi)$ is a trivial fixed point of the classical evolution equation \eqref{eq:eom_cooling} or \eqref{eq:eom_cont}.
In fact, this state corresponds to the analytically constructed nonequilibrium stationary state for the engineered dissipative process under consideration \cite{Diehl:2008}.
Using the approximate treatment of the TWA, however, this fixed point solution is not unique. 
For instance, for a one-dimensional system with periodic boundary conditions, there exists a family of solutions with constant amplitude $\Phi_0$ that is given by
\begin{equation}
 \lim_{t\to\infty}\varphi_j(t)=\Phi_0 e^{i\Theta_{j,l}} \quad {\rm with} \quad \Theta_{j,l}=\Theta_0+\frac{2\pi j l}{N} \ .
\end{equation}
The integer $l\in\{0,N-1\}$ determines how often the phase winds around the circle.
In fact, each ensemble member ends up in one of these different winding number sectors depending on the initial configuration $\bs\varphi_0$.

Unfortunately, we have not yet found a full characterization of the possible fixed points of the nonlinear field equation \eqref{eq:eom_cooling} in more than one dimension.
However, this observation indicates that the semiclassical analysis using the TWA does not capture the full quantum dynamics of dissipative Bose-Einstein condensation, for which the analytically constructed nonequilibrium stationary state is unique.
In fact, this is not completely unexpected as it is well known that the TWA is only asymptotically exact at short times but breaks down at late times due to the absence of genuine quantum fluctuations \cite{Polkovnikov:2009ys}.
It is, however, still an interesting question how much of the dynamics of dissipative Bose-Einstein condensation is captured by a semiclassical analysis using the TWA in order to provide insight into necessary extensions of the semiclassical analysis.

To study the real-time dynamics of the lattice boson system numerically, we solve the classical equations of motion \eqref{eq:eom_cooling} and compute observables according to \eqref{eq:obs}.
Therefore, we have to specify an initial density matrix $\rho(t_0)$ or, equivalently, the Wigner function $\rho_W(\bs\varphi_0,t_0)$ from which the initial configurations $\bs\varphi_0$ are sampled.
We note that the Wigner function is not positive definite in general.
Since Monte Carlo importance sampling requires a positive definite integration measure, the choice of viable initial density matrices is restricted.
To study the generation of phase coherence via engineered dissipation, we assume that the the bosonic degrees of freedom at the individual lattice sites are completely uncorrelated at initial time $t_0$.
Moreover, we assume that the average occupation at each lattice site is given by $\langle a_i^\dagger a_i\rangle(t_0)=N_0\gg1$.
These assumptions are fulfilled by the coherent state density matrix
\begin{equation}
 \rho(t_0)=e^{-|\bs\psi|^2}\ket{\bs\psi}\bra{\bs\psi} \ ,
\end{equation}
with $\psi_i=\sqrt{N_0}e^{i\theta_i}$.
Due to the fact that the individual lattice sites are supposed to be uncorrelated at initial time $t_0$, we average over all possible phases $\theta_i\in[0,2\pi)$.
Accordingly, we obtain a positive definite Wigner function
\begin{equation}
 \label{eq:wigner_init}
 \rho_W(\bs{\varphi}_0,t_0)=\frac{1}{\pi^{N}}\prod_{i=1}^{N}\int_{0}^{2\pi}d\theta_i\exp(-2|\varphi_{i}(t_0)-\sqrt{N_0}e^{i\theta_i}|^2) \ .
\end{equation}

\begin{figure}[h!]
\includegraphics[width=0.97\columnwidth]{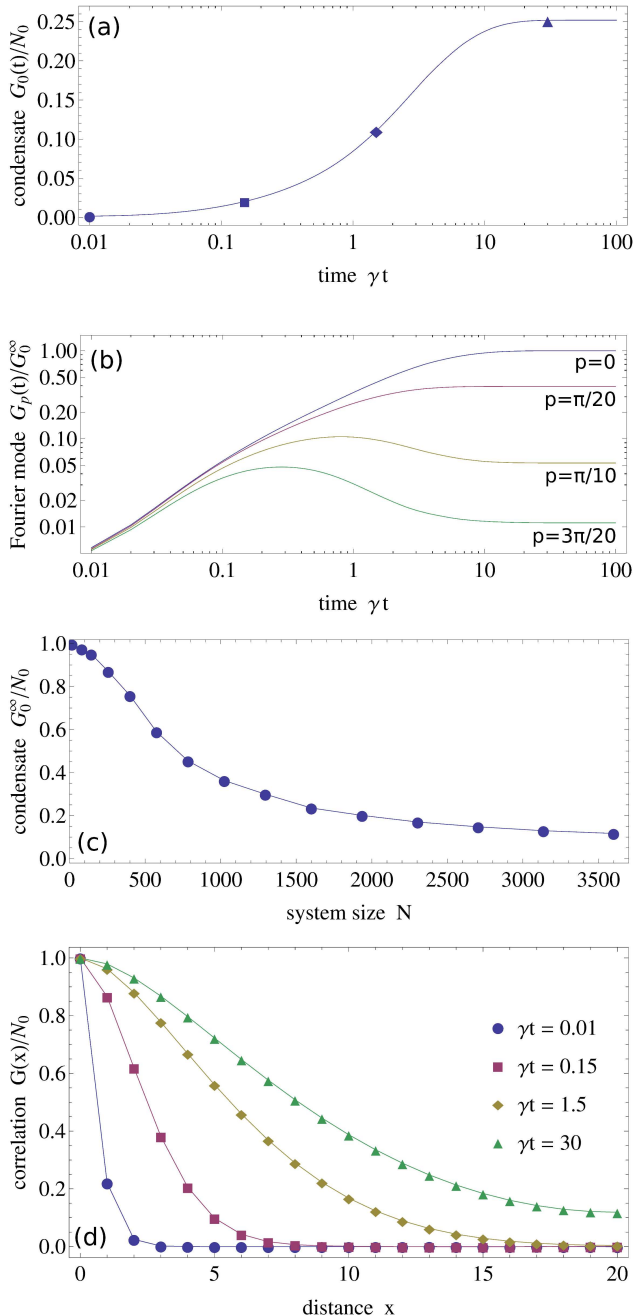}
\caption{[\textit{Color online}]
{\it (a)} Time evolution of the condensate mode $G_0(t)$ for $N=40^2$ (logarithmic abscissa). The asymptotic value $G_0^{\infty}\sim N_0/4<N_0$ shows that no global phase coherence is established. 
{\it (b)} Time evolution of some low-lying Fourier modes $G_p(t)$ for different momenta $p_1=\{0,\pi/20,\pi/10,3\pi/20\}$ and $p_2=0$ (double-logarithmic scale). 
{\it (c)} Asymptotic value $G_0^{\infty}$ as a function of the lattice size $N$. While there is still global phase coherence for $N=4^2$, less coherence is generated on ever larger systems.
{\it (d)} To illustrate the growth of correlations we display the correlation function $G(|\bs x_i-\bs x_j|,t)$ at different times, corresponding to the instants indicated by the markers in (a). 
\label{fig1:condensate}}
\end{figure}

\begin{figure*}[t!]
 \includegraphics[width=1.97\columnwidth]{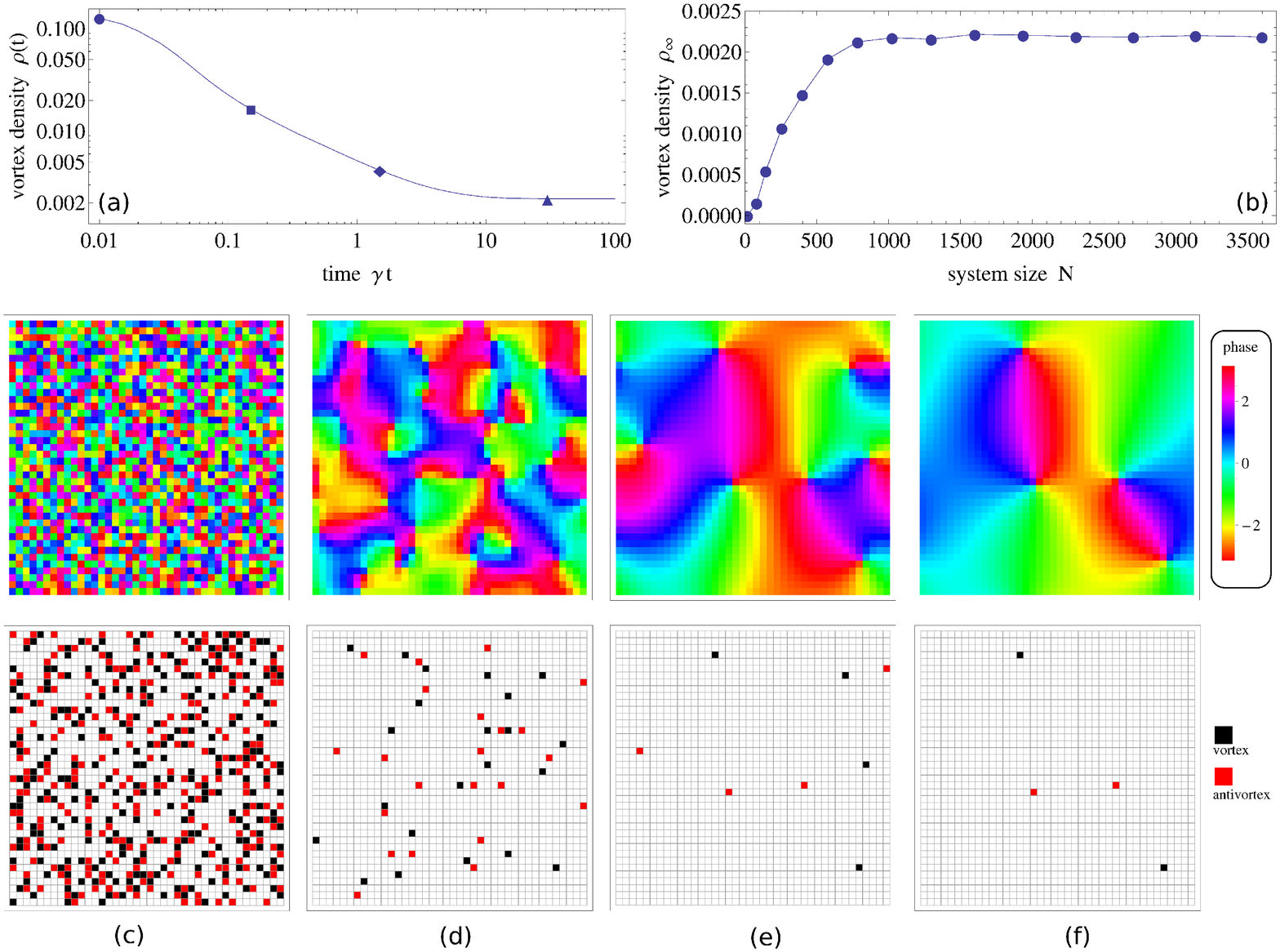}
 \caption{[\textit{Color online}]
 {\it (a)} Time evolution of the vortex density $\rho(t)$ for $N=40^2$ (double-logarithmic scale). The markers denote the times at which the snapshots are shown.
 {\it (b)} Asymptotic vortex density $\rho_{\infty}$ as a function of the lattice size $N$. In lattice units, the vortex density saturates at $\rho_\infty= 2.18(4)\cdot 10^{-3}$ for $N\gtrsim32^2$.
 {\it (c)--(f)} Snapshots of the local bosonic phases (upper) and the vortex-antivortex distribution (lower) for a random realization at different times (c) $t=0$, (d) $\gamma t=0.15$, (e) $\gamma t=1.5$ and (f) $\gamma t=30$.
 \label{fig2:vortices}}
\end{figure*}

\subsection{Generation of phase coherence}         

In the following we consider a bosonic system on a square lattice with periodic boundary conditions and an initial average occupation $N_0=10$.
To study the growth of phase coherence in time, we define the correlation function
\begin{equation}
 G(|{\bs x}_i-{\bs x}_j|,t)=\langle a_i^\dagger a_j\rangle(t) \ ,
\end{equation}
where we used spatial translation invariance.
Hence, we may also characterize the real-time evolution in terms of Fourier modes
\begin{equation}
 G_{p}(t)=\frac{1}{N^2}\sum_{i,j=1}^{N}e^{i\bs{p}\cdot(\bs x_i-\bs x_j)}G(|\bs x_i-\bs x_j|,t) \ ,
\end{equation}
with $p_i=2\pi q_i/N$ and $q_i\in\{0,\ldots,N-1\}$.
The Fourier component with zero momentum, $p=(p_1,p_2)=0$, is denotes as the condensate mode $G_0(t)$.

In Fig.~\ref{fig1:condensate}a we display the time evolution of the condensate mode mode $G_0(t)$ for $N=40^2$ lattice sites for an ensemble size of $N_{\rm samp}=3200$ configurations.
We observe that a finite condensate is quickly established due to the dissipative cooling mechanism.
However, the reached asymptotic value is below the maximum possible value
\begin{equation}
 \lim_{t\to\infty}G_{0}(t)=G_0^{\infty}\simeq N_0/4<N_0 \ ,
\end{equation}
which would correspond to global phase coherence.
The time evolution of the low-lying Fourier modes is displayed in Fig.~\ref{fig1:condensate}b.
The action of the quasi-local quantum jump operators gradually correlates bosons over ever larger distances.
We emphasize, however, that the low-lying Fourier modes have a sizable amplitude at late times which again indicates that no global phase coherence is established.
We note that this is in contract to dissipative Bose-Einstein condensation of hardcore bosons for which all low-lying Fourier modes decay at late times \cite{Caspar:2015,Caspar:2016}.

In fact, the asymptotic value $G_{0}^{\infty}$ strongly depends on the lattice size $N$ as shown in Fig.~\ref{fig1:condensate}c.
While there is still global phase coherence for $N=4^2$, less coherence is generated for ever larger systems. 
Finally, in Fig.~\ref{fig1:condensate}d we again support this observation by showing the correlation function $G(|\bs x_i-\bs x_j|,t)$ at different instants of time.
We find that the correlation grows at all distances, however, it exhibits a non-exponential decay at large distances and asymptotic times.

The discrepancy between our results and the analytically constructed nonequilibrium stationary state is traced back to the approximate treatment within the TWA.
We emphasize, however, that while the TWA does not capture the full quantum dynamics it already accounts for a large part of the dissipative condensation phenomenon:
The classical equations of motion \eqref{eq:eom_cooling} describe the generation of phase coherence over large distances.
However, the TWA does not correctly describe the global phase coherence at late times as the equations of motion do not have a unique fixed point, as discussed above.
To remedy this shortcoming of the TWA, it will be necessary to account for genuine quantum fluctuations to the TWA by taking into account higher order terms in the response field $\bs\eta_k$ \cite{Polkovnikov:2003}, cf.~Section~\ref{sec:twa}.
The systematic inclusion of quantum fluctuations beyond the TWA for open quantum systems is beyond the scope of the current work but will be investigated in future research.

\subsection{Dynamics of vortex-antivortex pairs}

In the following, we show that our TWA results can be conveniently described in terms of vortices.
To this end, we define the vortex density
\begin{equation}
 \rho(t)=\frac{1}{N}\langle N_{V}(t)\rangle=\frac{1}{N}\langle N_{A}(t)\rangle \ ,
\end{equation}
where $N_{V}(t)$ denotes the number of vortices which equals the number of antivortices, $N_A(t)$.
Using the density--phase representation for the bosonic degrees of freedom, $\varphi_i(t)=\sigma_i(t) e^{i\theta_i(t)}$, the vortices (antivortices) are detected by a positive (negative) winding of the bosonic phases $\theta_i$ around a fundamental plaquette.
For later use, we also define the vortex-antivortex configuration $\bs{\xi}(t)$, where $\xi_i=\pm1$ at the position of the vortex (antivortex) and zero elsewhere.

In fact, the initial Wigner function \eqref{eq:wigner_init} is characterized by a completely random distribution of the bosonic phases, which can be regarded as a vortex-antivortex condensate at high temperatures, cf.~Fig.~\ref{fig2:vortices}c.
Upon applying the quantum jump operators \eqref{eq:jump}, these vortices become mobile so that vortex-antivortex pairs annihilate and phase coherence over ever larger regions is established, cf.~Fig.~\ref{fig2:vortices}d-e.
We emphasize, however, that not all vortices are annihilated in general so that the final state of the time evolution is characterized by a finite number of frozen vortices, cf.~Fig.~\ref{fig2:vortices}f.
In Fig.~\ref{fig2:vortices}b we display the asymptotic value of the vortex density $\lim_{t\to\infty}\rho(t)=\rho_{\infty}$ as a function of the lattice size $N$.
While all vortices are annihilated for small systems $N=4^2$, a finite number of vortices remains for ever larger lattices.
The vortex density saturates at $\rho_\infty= 2.18(4)\cdot 10^{-3}$ for lattices as large as $N\gtrsim 32^2$.
It is again an artifact of the approximate treatment within the TWA that a finite number of vortex-antivortex pairs survives the dissipative cooling process.

\begin{figure}[t!]
\includegraphics[width=\columnwidth]{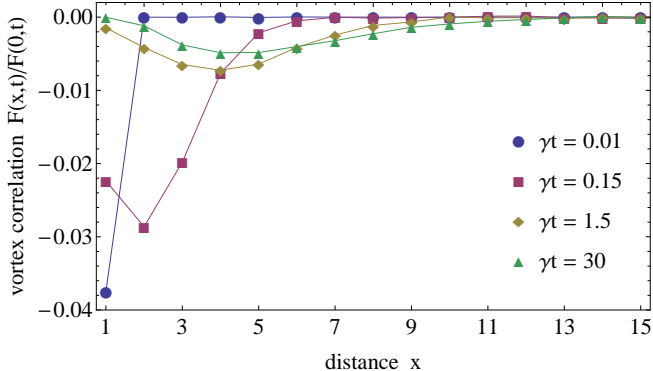}
\caption{[\textit{Color online}]
 Vortex-vortex correlation function $F(x,t)$ for $N=40^2$ at different times as indicated by the markers in Fig.~\ref{fig2:vortices}a.
 The curve for $\gamma t=0.01$ is rescaled by a factor of $0.15$ to fit into the figure.
\label{fig3:vortex_correlation}}
\end{figure}

To obtain further information on the vortex structure we study the vortex-vortex correlation function
\begin{equation}
 F(|\bs x_i-\bs x_j|,t)=\frac{1}{N}\langle \xi_i(t) \xi_j(t) \rangle 
\end{equation}
with $F(0,t)=2\rho(t)$ \cite{Liu:1992}, whose minimum indicates the preferred distance between vortices and antivortices.
In Fig.~\ref{fig3:vortex_correlation} we display $F(|\bs x_i-\bs x_j|,t)$ for $N=40^2$ lattice sites for an ensemble size of $N_{{\rm samp}}=31200$ at different instants of time.
We find that nearest-neighbor vortices have on average an opposite sign.
While vortex-antivortex pairs dominantly reside on nearest-neighbor sites in the vortex condensate phase at $t=0$, the average distance $d(t)$ between vortices and antivortices increases as time evolves.
At asymptotic times we obtain $\lim_{t\to\infty}d(t)\sim4-5$ lattice sites, which is independent of the lattice size for $N\gtrsim 32^2$, i.e., in the regime where the vortex density saturates, cf.~Fig.~\ref{fig2:vortices}b.

The exact nonequilibrium stationary state is a Bose-Einstein condensate by construction \cite{Diehl:2008}, i.e., it is free of vortex-antivortex pairs.
However, our analysis within the TWA yields that a finite number of frozen vortex-antivortex pairs survives the dissipative cooling process.
This behavior can be interpreted as follows:
In the TWA, the classical equations of motion \eqref{eq:eom_cooling} appropriately describes the early-time annihilation of vortex-antivortex pairs starting from the vortex condensate phase.
However, as soon as either of the fixed points of the classical evolution equation is approached, the vortices that are then still present are distributed in such a way that they are on average $4-5$ lattice sites apart from each other and freeze.
Without including further quantum corrections, these frozen vortex-antivortex pairs describe the asymptotic state within the TWA.
Accordingly, quantum fluctuations beyond the TWA, that exert kicks on the remaining vortices and induce ongoing dynamics, will be essential to correctly describe the process of dissipative Bose-Einstein condensation in real time.
\newpage

\section{Conclusions \& outlook}
\label{sec:Conclusions}

For the first time we studied the initial value problem for a large driven-dissipative lattice boson systems in two spatial dimensions that is governed by a Markovian quantum master equation.
To this end, we employed the Wigner-Weyl phase space quantization and derived an exact functional integral expression for the time-evolved Wigner function \eqref{eq:wig_evolved}.
Since it is in general not possible to evaluate this expression analytically, we employed the truncated Wigner approximation (TWA) in which quantum fluctuations are only taken into account in the initial state whereas the dynamics is governed by classical evolution equations \eqref{eq:eom}.

This method was then used to study the real-time dynamics of an engineered dissipative process that drives the system of lattice bosons into a nonequilibrium stationary state with global phase coherence (Bose-Einstein condensate) \cite{Diehl:2008}.
Using numerical simulation, we found that the TWA does not fully capture the dynamics of the dissipative cooling process:
On the one hand, the TWA correctly describes that the application of the quantum jump operators \eqref{eq:jump} generates long-range correlations in the system.
On the other hand, the TWA does not yield the global phase coherent state as unique fixed point of the time evolution.

We explained this behavior by adopting the vortex picture:
The initial state corresponds to a vortex condensate and the dissipative process causes vortex-antivortex pairs to annihilate.
The vortex-antivortex annihilation corresponds to the dissipative cooling of the bosonic system.
Given a random initial configuration $\bs\varphi_0$, the final state $\lim_{t\to\infty}\bs\varphi(t)$ contains a finite number of vortex-antivortex pairs in general.
Within the TWA, these vortices are distributed in such a way that nearest-neighbor vortices have on average an opposite sign and are on average $4-5$ lattice sites apart from each other.
This behavior can be traced back to the fact that the classical equations of motion \eqref{eq:eom_cooling} do not have a unique fixed point.

Two immediate questions arise which go beyond the scope of this work but will be investigated in future research.
First, we have seen that the final state of the classical time evolution is characterized by a nonvanishing vortex density $\rho_{\infty}$, for which the individual vortex configurations $\lim_{t\to\infty}\bs{\xi}(t)$ are frozen in each realization.
In the full quantum theory, however, we expect quantum fluctuations to induce ongoing dynamics so that the remaining vortex-antivortex pairs annihilate. 
In order to study the full quantum dynamics, however, it will be necessary to systematically extend the TWA, which only accounts for initial state quantum fluctuations.
In principle, this can be done by performing a higher order expansion of the exact expression for the time-evolved Wigner function \eqref{eq:wig_evolved} with respect to the response field $\bs\eta_k$.
The corresponding quantum corrections can then be implemented either as a nonlinear response of observables or in terms of stochastic quantum jumps \cite{Polkovnikov:2003}.
This future investigation will allow us to determine the time scales at which the Bose-Einstein condensate is formed and to make clear statements on the efficiency of this specific state preparation protocol. 

Second, we found that the asymptotic density of vortex-antivortex pairs $\rho_\infty$ is a unique number, i.e., it does not depend on the value of the rate $\gamma$.
This is due to the fact that the system does not contain an explicit time scale as $\gamma$ can be scaled out.
Besides extending the TWA to account for quantum fluctuations, it will be most interesting to see how the inclusion of coherent Hamiltonian dynamics, which introduces another energy or time scale, alters the dissipative vortex dynamics.
For instance, it is still an open question whether the dissipative cooling through a critical point of, e.g., the Bose-Hubbard Hamiltonian, results in the same universal behavior of topological objects as if the system is quenched through the critical point (dissipative Kibble-Zurek mechanism). 
The application of versatile functional integral methods will give us unprecedented insight into the role of topological objects for the dynamics of large driven-dissipative quantum many-body systems in more than one spatial dimension.

\subsection*{Acknowledgments}
We thank S.~Diehl, V.~Kasper, D.~Mesterh\'{a}zy and U.-J.~Wiese for discussions.
This research is funded by the European Research Council under the European Union’s Seventh Framework Programme (FP7/2007-2013)/ERC under grant agreement 339220.

\end{document}